\documentclass[12pt,a4paper]{article}
\usepackage{graphicx}
\usepackage[T1]{fontenc}
\usepackage[utf8]{inputenc}
\usepackage{textcomp}
\usepackage[sc,osf]{mathpazo}
\usepackage{a4wide}  
\usepackage{latexsym,amsthm,amsfonts,amsmath,mathrsfs,amssymb}
\usepackage{dsfont}
\usepackage{accents}
\usepackage[nosort]{cite}
\usepackage{booktabs} 
\usepackage[unicode,implicit]{hyperref}
\hypersetup{%
  pdftitle    = {On-shell Lagrangians as total derivatives and the generalized
    Komar charge}
  pdfkeywords = {conserved quantities, generalized Komar charge, Euler
    theorem, global symmetries},
  pdfauthor   = {Jos\'e Luis V. Cerdeira and Tom\'as Ort\'{\i}n},
  plainpages  = true,
  colorlinks  = true,
  citecolor   = blue,
  urlcolor    = red,
  linkcolor   = black
}
\newcommand{\hepth}[1]{{\tt
\href{http://www.arXiv.org/abs/hep-th/#1}{hep-th/#1}}}

\newcommand{\arxiv}[1]{{\tt arXiv:\href{http://www.arXiv.org/abs/#1}{#1}}}
\newcommand{\doi}[1]{{\tt DOI:\href{http://dx.doi.org/#1}{#1}}}

\allowdisplaybreaks

\makeatletter
\@addtoreset{equation}{section}
\makeatother

\begin{document}

\begin{flushright}
\small
IFT-UAM/CSIC-25-064\\
June 17\textsuperscript{th}, 2025\\
\normalsize
\end{flushright}

\vspace{.2cm}

\begin{center}

  {\Large {\bf On-shell Lagrangians as total derivatives\\[.5cm] and the
      generalized Komar charge}}
 
\vspace{1cm}

\renewcommand{\thefootnote}{\alph{footnote}}
{\sl Jos\'e Luis V.~Cerdeira}$^{1,2,}$\footnote{Email: {\tt jose.verez-fraguela[at]estudiante.uam.es}}
{\sl and Tom\'{a}s Ort\'{\i}n}$^{2,}$\footnote{Email: {\tt Tomas.Ortin[at]csic.es}}

\setcounter{footnote}{0}
\renewcommand{\thefootnote}{\arabic{footnote}}
\vspace{1cm}

${}^{1}${\it\small Instituto de F\'{\i}sica Corpuscular (IFIC), University of
  Valencia-CSIC,\\
Parc Cient\'{\i}fic UV, C/ Catedr\'atico Jos\'e Beltr\'an 2, E-46980 Paterna, Spain}

\vspace{0.2cm}

${}^{2}${\it\small Instituto de F\'{\i}sica Te\'orica UAM/CSIC\\
C/ Nicol\'as Cabrera, 13--15,  C.U.~Cantoblanco, E-28049 Madrid, Spain}

\vspace{3cm}

{\bf Abstract}
\end{center}
\begin{quotation}
  {\small Lagrangians which transform homogeneously under a global
    transformation of the fields (a global rescaling, for instance) can be
    written on-shell as a total derivative which has a universal,
    solution-independent expression, using a functional version of the Euler
    theorem for homogeneous functions.  We study the uniqueness of this
    expression and how this result can be used in the construction of
    generalized Komar charges.}
\end{quotation}

\newpage
\pagestyle{plain}

\section{Introduction}

Charge conservation is one of the most powerful tools one can use to
characterize the state of a system and predict its evolution. The
identification and definition of the conserved charges of a system or theory,
tied to the identification of its symmetries by Noether's theorems, is one of
the most fundamental tasks to be performed in its study. In the context of
theories that describe gravity, conserved charges are defined as ``surface''
integrals\footnote{By ``surface integrals'' we mean integrals over spacelike
  $(d-2)$-dimensional submanifolds of a $d$-dimensional spacetime.} at
infinity of $(d-2)$-form ``charges'' associated to a generator of general
coordinate transformations (GCTs) (a vector field). These $(d-2)$-form charges
are asymptotically closed on-shell when the vector field is asymptotically
Killing, as in the Abbott--Deser construction
\cite{Abbott:1981ff}\footnote{See also
  Refs.~\cite{Barnich:2001jy,Deser:2002jk} and the review
  ref.~\cite{Adami:2017phg} for more general and recent analyses.} or globally
closed on-shell when the vector field is a global Killing vector, as in the
Komar charge of pure gravity \cite{Komar:1958wp}.\footnote{See
  Ref.~\cite{Barnich:2003xg} for similar constructions in other theories with
  gauge symmetries} $(d-2)$-form charge which are globally closed on-shell
give rise to Gauss laws. In the case of the Komar $(d-2)$-form charge the
Gauss law leads to the Smarr formula that relates the thermodynamical
functions, charges and potentials of the stationary black-hole solutions of
General Relativity (GR) in vacuum (Kerr black holes)
\cite{Bardeen:1973gs,Carter:1973rla}. The Gauss law can also be used to
constrain the possible solutions of the theory

The Komar charge of GR in vacuum was generalized in
Refs.~\cite{Magnon:1985sc,Bazanski:1990qd} to include a cosmological constant
and the coupling to a Maxwell field. A general procedure to systematically
construct the (on-shell-closed) generalized Komar charges of more general
theories based on the observations of Ref.~\cite{Liberati:2015xcp} has been
developed in
Refs.~\cite{Ortin:2021ade,Mitsios:2021zrn,Meessen:2022hcg,Ortin:2022uxa,Ortin:2024mmg}
and it has been successfully applied in theories ranging from those describing
scalars coupled to gravity \cite{Ballesteros:2023muf,Ballesteros:2024prz} to
supergravity \cite{Bandos:2023zbs,Bandos:2024rit} and superstring effective
actions \cite{Ortin:2024emt}.

One of the key steps in the algorithm is the calculation of the on-shell
Lagrangian volume form $\mathbf{L}$,\footnote{We are going to use
  differential-form notation throughout the whole paper. Our conventions are
  those of Ref.~\cite{Ortin:2015hya}.} which has to be evaluated for the class
of solutions considered, \textit{i.e.}~those admitting a global Killing vector
$k$.\footnote{Notice that the calculation of the on-shell Lagrangian for
  solutions admitting a global Killing vector is essential in the Euclidean
  approach to quantum gravity applied to stationary black holes. Our results
  may also have applications in this context.} It can be shown (see
Section~\ref{sec-generalizedKomarcharges} for a more precise statement) that
the $(d-1)$-form $\imath_{k}\mathbf{L}_{\rm on-shell}$ is the total derivative
of a $(d-2)$-form $\boldsymbol{\omega}[k,\varphi]$ which is an essential
ingredient of the generalized Komar $(d-2)$-form (see
Section~\ref{sec-generalizedKomarcharges}). In any case, we need
$\boldsymbol{\omega}[k,\varphi]$ to have a ``universal form'' whose integral
for any solution admitting a Killing vector $k$ can be identified in terms of
conserved charges or their associated chemical potentials.

Writing $\mathbf{L}$ on-shell as a total derivative in a suitable form can
prove difficult and in this paper we are going to show how it can be done in a
systematic way when there is a global transformation of the fields whose net
effect is a rescaling of $\mathbf{L}$. The main, but not unique example, is
that of Lagrangians which are \textit{quasi-homogeneous} in the
fields.\footnote{The Lagrangian scales with a given weight when the fields are
  rescaled with weights that can be different for each of them.} In that case
our result takes the form of an Euler theorem in field theory.  While this
seems to exclude the important case of non-linear Lagrangians, we are going to
see that one can still find this kind of transformations if the coupling
constants are promoted to scalar fields constrained to be constant by a
Lagrange-multiplier term. Thus, we expect this method to be of wide
applicability and we are going to show in several examples how the results are
used to construct the generalized Komar $(d-2)$-form charge of the theory.

This paper is organized as follows: in Section~\ref{eq:Eulertheorem} we prove
our main result. In Section~\ref{sec-generalizedKomarcharges} we review the
construction of generalized Komar charges in general, GCT-invariant, field
theories. In Sections~\ref{sec-example1EMD}, \ref{sec-example2N1D5} and
\ref{sec-example3selfinteractingscalar} we show how to use our main result to
construct the generalized Komar charges of the Einstein--Maxwell--Dilaton
theory, of the bosonic sector of minimal 5-dimensional supergravity and of a
self-interacting scalar field coupled to gravity (which provides an example of
non-linear theory). Section~\ref{sec-conclusions} contains our conclusions.

\section{The Euler theorem for Lagrangians}
\label{eq:Eulertheorem}

Consider a Lagrangian $d$-form $\mathbf{L}$ which is a function of a number of
$p_{i}$-form fields $\varphi^{i}$, that we collectively denote by $\varphi$,
and their derivatives. Under an arbitrary, infinitesimal, variation of the
fields, the action

\begin{equation}
  \label{eq:genericaction}
  S[\varphi]
  \equiv
  \int \mathbf{L}\,,
\end{equation}

\noindent
changes as\footnote{Sums over repeated indices are understood.}

\begin{equation}
  \label{eq:genericvariationoftheaction}
  \delta S
  =
  \int \left\{ \mathbf{E}_{i}\wedge \delta\varphi^{i}
    +d\mathbf{\Theta}(\varphi,\delta\varphi) \right\}\,,
\end{equation}

\noindent
where

\begin{equation}
  \mathbf{E}_{i}
  \equiv
  \frac{\delta\mathbf{L}(\varphi)}{\delta\varphi}\,,
\end{equation}

\noindent
are the Euler--Lagrange equations of motion and
$\mathbf{\Theta}(\varphi,\delta\varphi)$ is the pre-symplectic potential,
which is linear in $\delta\varphi$.

For simplicity, we are going to start considering just scaling
transformations. Let us assume that, under a rescaling of the fields with
weights $\omega_{i}$, the Lagrangian rescales with weight $\omega_{L}$

\begin{equation}
  \mathbf{L}(\lambda^{\omega}\varphi)
  =
  \lambda^{\omega_{L}}\mathbf{L}(\varphi)\,.
\end{equation}

\noindent
Infinitesimally, we have

\begin{equation}
  \delta_{\lambda}\varphi^{i}
  =
  \omega_{(i)}\lambda\varphi^{(i)}\,,
  \hspace{1cm}
  \delta_{\lambda}\mathbf{L}
  =
  \omega_{L}\lambda\mathbf{L}\,,
\end{equation}

\noindent
and particularizing Eq.~(\ref{eq:genericvariationoftheaction}) for these
transformations and setting $\lambda=1$ we arrive to

\begin{equation}
  \omega_{i}\mathbf{E}_{i}\wedge \varphi^{i}
  +d \mathbf{\Theta}(\varphi,\omega\varphi)
  =
  \omega_{L}\mathbf{L}\,,
\end{equation}

\noindent
so that\footnote{We indicate with $\doteq$ identities which may only hold
  on-shell.}

\begin{equation}
  \mathbf{L}
  =
  \frac{\omega_{i}}{\omega_{L}}\mathbf{E}_{i}\wedge \varphi^{i}
  +d \mathbf{\Theta}\left(\varphi,\frac{\omega}{\omega_{L}}\varphi\right)
  \doteq
   d \mathbf{\Theta}\left(\varphi,\frac{\omega}{\omega_{L}}\varphi\right)\,.
\end{equation}

Thus, in order to evaluate the Lagrangian or the action on any solution of the
equations of motion $\mathbf{E}_{i}=0$ it is enough to evaluate the exterior
derivative of the pre-symplectic potential with $\delta\varphi^{i}$ replaced
by $\frac{\omega_{(i)}}{\omega_{L}}\varphi^{(i)}$. 

However, in general, there could be more than one rescaling of the fields
resulting in a rescaling of the Lagrangian with weight $\omega_{L}$. The
difference between any two such rescalings will be a rescaling of the fields
that leaves the Lagrangian invariant, that is, it will be a global scaling
symmetry of the theory. If we label all these rescalings with $n=1,\ldots$,
for each of them we will have

\begin{equation}
  d\mathbf{J}^{n}
  =
  \omega^{n}_{i}\mathbf{E}_{i}\wedge \varphi^{i}
  \doteq
  0\,,
\end{equation}

\noindent
where $\omega^{n}_{i}$ is the weight of the $i$th field under the $n$the
rescaling. $\mathbf{J}^{n}$ is the Noether $(d-1)$-form current associated to
the $n$th scaling symmetry.

Writing

\begin{equation}
  \omega^{0}_{i}
  \equiv
  \omega_{i}/\omega_{L}\,
  \,\,\,\,\,\,
  \text{and}
  \,\,\,\,\,\,
  \mathbf{\Theta}\left(\varphi,\frac{\omega}{\omega_{L}}\varphi\right)
  \equiv
  \mathbf{J}^{0}\,,
\end{equation}

\noindent
we find that the most general expression for the Lagrangian is a linear
combination with coefficients $\alpha_{n}$ ($\alpha_{0}=1$)

\begin{equation}
  \label{eq:Lasatotalderivative}
  \mathbf{L}
  =
  \alpha_{n}\omega^{n}_{i}\mathbf{E}_{i}\wedge \varphi^{i}
  +d  \left(\alpha_{n}\mathbf{J}^{n}\right)
  \doteq
  d  \left(\alpha_{n}\mathbf{J}^{n}\right)\,.
\end{equation}

Since $d\mathbf{J}^{n}\doteq 0$, we only need to consider
$\mathbf{J}^{0}$. However, the above expression tells us that $\mathbf{J}^{0}$
is not uniquely defined since one can always add to it linear combinations of
the $\mathbf{J}^{n}$, $n>1$, which are closed on-shell. The constants
$\alpha_{n}$ can be determined by our desire to obtain a particular result (a
particular conserved charge) when we integrate. In most cases (like the
examples considered in this paper) we will not need to have any $\alpha_{n}$
different from zero but there may be cases in which we have to combine
$\mathbf{J}^{0}$ with another conserved current \cite{Barbagallo:2025fkg}.

Before closing this section we notice that, a a matter of fact, if we are just
interested in the expression of the Lagrangian as a total derivative, one can
use any global transformation (not necessarily a rescaling of the fields) that
rescales the Lagrangian with a non-zero weight. $\mathbf{J}^{0}$ will always
be given in terms of the pre-symplectic potential, now evaluated for
variations of the fields $\delta\varphi$ which are not simply proportional to
the fields themselves. Furthermore, we can always combine $\mathbf{J}^{0}$
with any other conserved current, not just with the conserved current
associated to some scaling symmetry. An interesting example can be found in
Ref.~\cite{Barbagallo:2025fkg}.

\section{Generalized Komar charges }
\label{sec-generalizedKomarcharges}

Now, let us see how we can use the result of the previous section in the
calculation of generalized Komar charges.  First, let us review the its
definition and properties.

Let as assume that the generic action Eq.~(\ref{eq:genericaction}) is
invariant up to a total derivative under some local transformations of the
fields $\delta_{\xi}\varphi^{i}$

\begin{equation}
    \label{eq:gaugevariationoftheaction1}
  \delta_{\xi}S
  =
  -\int d\mathbf{B}[\xi,\varphi]\,.
\end{equation}

\noindent
Specializing Eq.~(\ref{eq:genericvariationoftheaction}) to this symmetry, we
find an alternative expression for this variation

\begin{equation}
  \label{eq:gaugevariationoftheaction2}
  \delta_{\xi} S
  =
  \int \left\{ \mathbf{E}_{i}\wedge \delta_{\xi}\varphi^{i}
    +d\mathbf{\Theta}(\varphi,\delta_{\xi}\varphi) \right\}\,,
\end{equation}

\noindent
and subtracting one from the other and taking into account that the result is
independent of the integration domain, we get

\begin{equation}
\mathbf{E}_{i}\wedge \delta_{\xi}\varphi^{i}
+d\left\{\mathbf{\Theta}(\varphi,\delta_{\xi}\varphi)
  +\mathbf{B}[\xi,\varphi] \right\}
=
0\,.
\end{equation}

This identity implies that the combination
$\mathbf{E}_{i}\wedge \delta_{\xi}\varphi^{i}$ is a total
derivative\footnote{In order to arrive at the total derivative, we must
  integrate by parts $\mathbf{E}_{i}\wedge \delta_{\xi}\varphi^{i}$ as many
  times as necessary to remove all the derivatives of the gauge parameters. In
  so doing we generate two terms: a total derivative,
  $d\mathbf{S}[\xi,\varphi]$, which may contain derivatives of the gauge
  parameters and a term linear in the gauge parameters with coefficients which
  are combinations of the equations of motion and their derivatives. This last
  term linear in the gauge parameters must vanish identically for any values
  of the gauge parameters and, then, those coefficients must vanish
  identically (off-shell). The identities that state that those coefficients
  vanish off-shell are the \textit{Noether identities}. We are then left with
  $d\mathbf{S}[\xi,\varphi]$.}

\begin{equation}
  \label{eq:Sdef}
  \mathbf{E}_{i}\wedge \delta_{\xi}\varphi^{i}
  =
  d\mathbf{S}[\xi,\varphi]\,,
\end{equation}

\noindent
and, using this result in the previous equation, we arrive to the off-shell
identity

\begin{subequations}
  \begin{align}
    d\mathbf{J}[\xi]
    & =
      0\,,
    \\
    & \nonumber \\
    \mathbf{J}[\xi]
    & \equiv
      \mathbf{\Theta}(\varphi,\delta_{\xi}\varphi)
      +\mathbf{S}[\xi,\varphi]
      +\mathbf{B}[\xi,\varphi]\,,
  \end{align}
\end{subequations}

\noindent
which implies the local existence of a $(d-2)$-form charge $\mathbf{Q}[\xi]$
(the \textit{Noether charge} associated to the local symmetry) such that
\begin{equation}
  \label{eq:Noetherchargedefinition}
  \mathbf{J}[\xi]
  =
  d\mathbf{Q}[\xi]\,.
\end{equation}

In general, the Noether $(d-2)$-form charge is not closed. By definition,

\begin{equation}
  \label{eq:dQ}
  d\mathbf{Q}[\xi]
  =
  \mathbf{J}[\xi]
  =
      \mathbf{\Theta}(\varphi,\delta_{\xi}\varphi)
      +\mathbf{S}[\xi,\varphi]
      +\mathbf{B}[\xi,\varphi]\,.  
\end{equation}

\noindent
However, we notice that 

\begin{subequations}
  \begin{align}
    \mathbf{\Theta}(\varphi,\delta_{\xi}\varphi)
    & \,\d{=}\,
      0\,,
    \\
    & \nonumber \\
\mathbf{S}[\xi,\varphi] 
    & \doteq
      0\,,
    \\
    & \nonumber \\
    \mathbf{B}[\xi,\varphi]
    & \,\d{=}\,
    d \boldsymbol{\omega}(\xi,\varphi)\,,
  \end{align}
\end{subequations}

\noindent
where we denote with $\d{=}$ identities that only hold when $\xi$ is a
\textit{reducibility} of \textit{Killing parameter}
\cite{Barnich:2001jy,Barnich:2003xg}, which, by definition, satisfies

\begin{equation}
  \delta_{\xi}\varphi
  \,\d{=}\,
  0\,,
  \,\,\,\,\,
  \forall\varphi^{i}\,.
\end{equation}

\noindent
The first of these equations follows from the linearity of the pre-symplectic
potential on $\delta_{\xi}\varphi$. The second follows from the definition of
$\mathbf{S}[\xi,\varphi]$ Eq.~(\ref{eq:Sdef}). The third equation follows from
the exact invariance of the action under $\delta_{\xi}$ for Killing parameters
($d\mathbf{B}[\xi,\varphi]=0$).

Then, we can modify the Noether charge and define, for every Killing parameter
$\xi$, the \textit{generalized Komar charge}

\begin{equation}
  \label{eq:generalizedKomarcharge}
  \mathbf{K}[\xi]
\equiv
  -\mathbf{Q}[\xi]+\boldsymbol{\omega}(\xi,\varphi)\,,  
\end{equation}

\noindent
which, as a consequence of its definition and the above observations, is closed
on-shell:

\begin{equation}
  d\mathbf{K}[\xi]
  \doteqdot
  0\,.
\end{equation}

We are interested in theories which are invariant under general coordinate
transformations (GCTs), whose infinitesimal form $\delta_{\xi}$ is generated
by vector fields $\xi =\xi^{\mu}\partial_{\mu}$. On tensor fields these
transformations act as the Lie derivative with respect to $\xi$, that is,
$\delta_{\xi} = -\mathcal{L}_{\xi}$, and, if the Lagrangian is a $d$-form,

\begin{equation}
  \delta_{\xi}S
  =
  -\int \mathcal{L}_{\xi}\mathbf{L}
  =
  -\int d\imath_{\xi}\mathbf{L}\,.
\end{equation}

However, as explained, for instance, in
Refs.~\cite{Elgood:2020svt,Elgood:2020mdx,Elgood:2020nls}\footnote{See
  Ref.~\cite{Prabhu:2015vua} for a more rigorous treatment in the framework of
  principal bundles which, however, does not cover more general cases such as
  the gauge symmetries of $p$-form potentials.}, when GCTs act on fields which
have gauge freedoms, they induce ``compensating'' or ``induced'' gauge
transformations $\delta_{\Lambda_{\xi}}$ and, generically,\footnote{Here
  $\delta_{\Lambda}$ stands for all the gauge transformations and $\Lambda$
  represents their gauge parameters. The subindex $\xi$ in $\Lambda_{\xi}$
  indicates that these gauge parameters depend on $\xi$. (In general, these
  compensating gauge transformation parameters $\Lambda_{\xi}$ depend on the
  fields as well, but we do not indicate this dependence to simplify the
  notation.)  Examples of these compensating gauge parameters will be
  presented in the following sections.}

\begin{equation}
  \label{eq:deltaxigauge}
\delta_{\xi} = -\mathcal{L}_{\xi} +\delta_{\Lambda_{\xi}}\,.  
\end{equation}

\noindent
Thus, when the Lagrangian is invariant under gauge transformations up to a
total derivative, (due, for instance, to the presence of Chern--Simons terms)

\begin{equation}
  \delta_{\Lambda}S
  =
  -\int d\mathbf{B}(\Lambda,\varphi)\,,
\end{equation}

\noindent
we have to take into account this total derivative in the transformation under
GCTs:

\begin{equation}
  \label{eq:totalderivativeGCTs}
  \delta_{\xi}S
  =
  -\int \delta_{\xi}\mathbf{L}
  =
  -\int d\left[\imath_{\xi}\mathbf{L}+\mathbf{B}(\Lambda_{\xi},\varphi)\right]
  \equiv
  -\int d\mathbf{B}(\xi,\varphi)\,.
\end{equation}

The Killing parameters of GCTs in these theories are Killing vectors and,
therefore, we will denote them by $k$. It is still true that, when we restrict
ourselves to the cases in which all the fields of the theory are annihilated
by $\delta_{k}$ defined as in Eq.~(\ref{eq:deltaxigauge}), since this implies
$ \delta_{k}S=0$,

\begin{equation}
  \mathbf{B}(k,\varphi)
  =
  \imath_{k}\mathbf{L}+\mathbf{B}(\Lambda_{k},\varphi)
  \,\d{=}\,
  d\boldsymbol{\omega}(k,\varphi)\,,
\end{equation}

\noindent
for some $\boldsymbol{\omega}(k,\varphi)$ that we want to determine if we want
to construct the generalized Komar charge. Actually, since we have to cancel
$\mathbf{S}(k,\varphi)$ as well, we have to make the above computation
on-shell:

\begin{equation} 
  \imath_{k}\mathbf{L}+\mathbf{B}(\Lambda_{k},\varphi)
  \doteqdot
  d\boldsymbol{\omega}(k,\varphi)\,.
\end{equation}

\subsection{Exactly gauge-invariant Lagrangians}

For simplicity, let us first consider the case in which the Lagrangian is
exactly gauge invariant, $\mathbf{B}(\Lambda_{k},\varphi)=0$. According to
Eq.~(\ref{eq:dQ}) and (\ref{eq:totalderivativeGCTs}) this implies that

\begin{equation}
  \label{eq:dQ=ikL}
  d\mathbf{Q}[k]
  \doteqdot
  \imath_{k}\mathbf{L}\,.
\end{equation}

\noindent
According to the above discussion $\imath_{k}\mathbf{L}$ on-shell is a total
derivative $\boldsymbol{\omega}(k,\varphi)$. This is the content of
Eq.~(\ref{eq:dQ=ikL}), but if we set
$\boldsymbol{\omega}(k,\varphi)=\mathbf{Q}[k]$, we will obtain an identically
vanishing generalized Komar charge.

In Ref.~\cite{Ortin:2024mmg}, an algorithm that produces a
$\boldsymbol{\omega}(k,\varphi)\neq \mathbf{Q}[k]$ and a non-trivial
generalized Komar charge was proposed. The algorithm was based on the following
observation: the term in the right-hand side of Eq.~(\ref{eq:dQ=ikL}) has been
obtained by taking first the interior product $\imath_{k}\mathbf{L}$ and then
evaluating it over some solution $s$. Following Ref.~\cite{Ortin:2024mmg} this
can be expressed in the form

\begin{equation}
  \label{eq:dQ=OsikL}
  \mathcal{O}_{s}d\mathbf{Q}[k]
  \doteqdot
  \mathcal{O}_{s}\imath_{k}\mathbf{L}\,,  
\end{equation}

\noindent
where the effect of the ``on-shell operator'' $ \mathcal{O}_{s}$ is to
evaluate the expression in the right over the solution $s$.

We can use a different algorithm to compute $\imath_{k}\mathbf{L}$ on-shell,
though. We can first use the results of Section~\ref{eq:Eulertheorem} to put
$\mathbf{L}$ in the form Eq.~(\ref{eq:Lasatotalderivative}) and then we can
evaluate it over the solution $s$. We can indicate this operation by
$\mathcal{O}_{s}\mathbf{L}$. The result is a total derivative. Then, we can
take the interior product with $k$, $\imath_{k}\mathcal{O}_{s}\mathbf{L}$.
The result is completely equivalent (up to equtions of motion and total
derivatives) to the one obtained computing
$\mathcal{O}_{s}\imath_{k}\mathbf{L}$ ($\mathcal{O}_{s}d\mathbf{Q}[k]$,
according to Eq.~(\ref{eq:dQ=OsikL})). However, as shown in
Ref.~\cite{Ortin:2024mmg} and references therein, it is \textit{formally}
different

\begin{equation}
  \imath_{k}\mathcal{O}_{s}\mathbf{L}
  =
  d\boldsymbol{\omega}(k,s)\,.
\end{equation}

\noindent
Consistency demands that the difference between these two results,
$d\left\{\boldsymbol{\omega}(k,s)-\mathbf{Q}[k]\right\}$ must vanish
on-shell. Since $\boldsymbol{\omega}(k,s)-\mathbf{Q}[k]$ evaluated on-shell is
nothing but the generalized Komar charge, according to the definition
Eq.~(\ref{eq:generalizedKomarcharge}), this proves 

\begin{equation}
  d\mathbf{K}[k]
  \doteqdot
  0\,.
\end{equation}

Thus, we can write \cite{Ortin:2024mmg}

\begin{equation}
  \label{eq:dKcommutator}
  d\mathbf{K}[k]
  =
  d\boldsymbol{\omega}(k,s)-\mathcal{O}_{s}d\mathbf{Q}[k]
  =
  \imath_{k}\mathcal{O}_{s}\mathbf{L}
  -\mathcal{O}_{s}\imath_{k}\mathbf{L}
  =
  [\imath_{k},\mathcal{O}_{s}]\mathbf{L}\,,
\end{equation}

\noindent
which should be understood as an algorithm to find $ d\mathbf{K}[k]$ and,
hence, $\mathbf{K}[k]$. It should also be understood that
$\mathcal{O}_{s}\mathbf{L}$ must be computed using
Eq.~(\ref{eq:Lasatotalderivative}). $\mathcal{O}_{s}\imath_{k}\mathbf{L}$
($\mathbf{Q}[k]$) can be computed using the standard methods, but, often, it
is more efficient to compute directly $\mathcal{O}_{s}\imath_{k}\mathbf{L}$.

Now, having computed $\mathcal{O}_{s}\mathbf{L}$ as a total derivative using
Eq.~(\ref{eq:Lasatotalderivative}), we must explain how to find
$\boldsymbol{\omega}(k,s)$.  Since we are assuming that $\delta_{k}$
annihilates all fields

\begin{equation}
  \label{eq:ikOsL}
  \imath_{k}\mathcal{O}_{s}\mathbf{L}
  \doteq
  \imath_{k}d\mathbf{J}^{0}
  \doteqdot
  -d\imath_{k}\mathbf{J}^{0} +\delta_{\Lambda_{k}}\mathbf{J}^{0}\,.
\end{equation}

\noindent
The second term must be a total derivative $d\boldsymbol{\tau} (k,s)$ if
$\mathbf{L}$ is exactly gauge-invariant and, therefore,

\begin{subequations}
  \begin{align}
    \boldsymbol{\omega}(k,s)
    & =
      -\imath_{k}\mathbf{J}^{0} +\boldsymbol{\tau} (k,s)\,,
    \\
\label{eq:generalizedKomarchargegeneral}
    \mathbf{K}[k]
    & =
      -\imath_{k}\mathbf{J}^{0} +\boldsymbol{\tau} (k,s)
      -\mathcal{O}_{s}\mathbf{Q}[k]
      =
      0\,,
  \end{align}
\end{subequations}

\noindent
where the Noether--Wald charge $\mathbf{Q}[k]$ can be computed using
Eq.~(\ref{eq:dQ=OsikL}).

\subsection{Gauge-invariant up to a total derivative Lagrangians}

If the Lagrangian is not exactly gauge invariant,
$\mathbf{B}(\Lambda_{k},\varphi)\neq 0$, and, according to Eq.~(\ref{eq:dQ})
and (\ref{eq:totalderivativeGCTs}),

\begin{equation}
  \label{eq:dQ=ikL+B}
\mathcal{O}_{s} d\mathbf{Q}[k]
  \doteqdot
  \mathcal{O}_{s}\imath_{k}\mathbf{L}
  +\mathcal{O}_{s}\mathbf{B}(\Lambda_{k},\varphi)\,.
\end{equation}

Now, in order to find an on-shell closed form, we have to express
$\imath_{k}\mathbf{L} +\mathbf{B}(\Lambda_{k},\varphi)$ on-shell as a total
derivative different from $\mathbf{Q}[k]$.\footnote{Eq.~(\ref{eq:dQ=ikL+B})
  already implies that, on-shell,
  $\imath_{k}\mathbf{L} +\mathbf{B}(\Lambda_{k},\varphi)$ is a total
  derivative ($d\mathbf{Q}[k]$ on-shell). We stress that we need to express
  this total derivative in a different form.} The only alternative is

\begin{equation}
  \imath_{k}\mathcal{O}_{s}\mathbf{L}
  +\mathcal{O}_{s}\mathbf{B}(\Lambda_{k},\varphi)\,,
\end{equation}

\noindent
and it leads, again, to Eq.~(\ref{eq:dKcommutator}). 

The first term is still given by Eq.~(\ref{eq:ikOsL}) while the second must
satisfy

\begin{equation}
  -\mathcal{O}_{s}d\mathbf{B}(\Lambda_{k},\varphi)
  =
  d\delta_{\Lambda_{k}}\mathbf{J}^{0}\,.
\end{equation}

\noindent
Then, we can write

\begin{equation}
  \delta_{\Lambda_{k}}\mathbf{J}^{0}
  =
  -\mathcal{O}_{s}\mathbf{B}(\Lambda_{k},\varphi)
  +d\boldsymbol{\tau}(k,s)\,,
\end{equation}

\noindent
for some $\boldsymbol{\tau}(k,s)$, and we recover the expression
Eq.~(\ref{eq:generalizedKomarchargegeneral}) for the generalized Komar charge.

Let us see  how to use this result in several examples.

\section{Example 1:  Einstein--Maxwell--Dilaton (EMD) gravity}
\label{sec-example1EMD}

The 4-dimensional EMD action is 

\begin{equation}
  \label{eq:actiondilatonBHs}
  \begin{aligned}
    S[e,A,\phi]
     & =
      \frac{1}{16\pi G_{N}^{(4)}}
      \int \left\{ -\star(e^{a}\wedge e^{b})
      \wedge R_{ab}
      +\tfrac{1}{2}d\phi\wedge \star d\phi
       +\tfrac{1}{2}e^{-a\phi}F\wedge \star F \right\}
    \\
     & \\
     & \equiv
       \int \mathbf{L}\,.
  \end{aligned}
\end{equation}

\noindent
Ignoring the overall factor $(16\pi G_{N}^{(4)})^{-1}$, the equations of
motion and pre-symplectic potential that follow from this action are

\begin{subequations}
  \begin{align}
    \mathbf{E}_{a}
    & =
      \imath_{a}\star(e^{b}\wedge e^{c})\wedge R_{bc}
      +\tfrac{1}{2}\left(\imath_{a}d\phi\wedge \star d\phi
      +d\phi\wedge \imath_{a}\star d\phi \right)
      \nonumber \\
    & \nonumber \\
    & \hspace{.5cm}
      +\tfrac{1}{2}e^{-a\phi}\left(\imath_{a}F\wedge \star F
      -F\wedge \imath_{a}\star F \right)\,,
    \\
    & \nonumber \\
    \label{eq:Ephi}
    \mathbf{E}_{\phi}
    & =
      -d\star d\phi
      -\tfrac{a}{2} e^{-a\phi}F\wedge \star F\,, 
    \\
    & \nonumber \\
        \label{eq:EA}
    \mathbf{E}_{A}
    & =
      -d\left(e^{-a\phi}\star F\right)\,,
    \\
    & \nonumber \\
      \label{eq:Theta}
    \mathbf{\Theta}(\varphi,\delta \varphi)
    & =
    -\star(e^{a}\wedge
    e^{b})\wedge \delta \omega_{ab} +e^{-a\phi} \star F\wedge\delta A
    +\star d\phi\delta\phi\,.
  \end{align}
\end{subequations}

The Noether--Wald charge of this theory can be obtained following
Refs.~\cite{Elgood:2020svt,Elgood:2020mdx,Elgood:2020nls,Ortin:2022uxa} and it
is given by \cite{Ballesteros:2023iqb}

\begin{equation}
  \mathbf{Q}[\xi]
  =
  \star (e^{a}\wedge e^{b}) P_{k\, ab} -P_{k}e^{-a\phi}\star F\,.
\end{equation}

In this expression $P_{k}$ is the Maxwell \textit{momentum map}, defined by
the  \textit{Maxwell momentum equation}

\begin{equation}
  \label{eq:Maxwellmomentummapequation}
  \imath_{k}F+dP_{k}
  =
  0\,,
\end{equation}

\noindent
and 
$P_{k\, ab}$ is the Lorentz \textit{momentum map}, defined by
the  \textit{Lorentz momentum equation}

\noindent
\begin{equation}
  \label{eq:Lorentzmomentummapequation}
  \imath_{k}R_{ab}+\mathcal{D}P_{k\, ab}
  =
  0\,.
\end{equation}

\noindent
This equation is always solved, at least, by the \textit{Killing bivector}

\begin{equation}
  \label{eq:Killingbivector}
  P_{k\, ab}
  =
  \nabla_{a}k_{b}\,.
\end{equation}

Under the global rescaling

\begin{equation}
  e^{\prime\, a}  =  \lambda e^{a}\,,
  \hspace{1cm}
  A' = \lambda A\,,
\end{equation}

\noindent
the action rescales with weight $2$. Thus, we can write

\begin{equation}
  \mathbf{L}
  =
  \tfrac{1}{2}\mathbf{E}_{a}\wedge e^{a}
  +\tfrac{1}{2}\mathbf{E}_{A}\wedge A
  +d\mathbf{J}^{0}\,.
\end{equation}

\noindent
with\footnote{$\omega^{ab}$ rescales with weight zero.}

\begin{equation}
  \mathbf{J}^{0}
  =
    \tfrac{1}{2}e^{-a\phi} \star F\wedge A\,.
\end{equation}

These theories are invariant under the global transformations

\begin{equation}
  e^{\phi'} = \lambda e^{\phi}\,,
  \hspace{1cm}
  A' = \lambda^{a/2}A\,,
\end{equation}

\noindent
leading to the identity

\begin{equation}
  d\mathbf{J}^{1}
  +\mathbf{E}_{\phi}\phi
  +\frac{a}{2}\mathbf{E}_{A}\wedge A
  =
  0\,,
\end{equation}

\noindent
with

\begin{equation}
  \mathbf{J}^{1}
  \equiv
  \star d\phi +\frac{a}{2}e^{-a\phi} \star F\wedge A\,,
\end{equation}

\noindent
and we can write the general expression

\begin{equation}
  \mathbf{L}
  =
  \tfrac{1}{2}\mathbf{E}_{a}\wedge e^{a}
  +\alpha\mathbf{E}_{\phi}\phi
  +\tfrac{1}{2}\left(1+a\alpha\right)\mathbf{E}_{A}\wedge A
  +d\left(\mathbf{J}^{0}+\alpha\mathbf{J}^{1}\right)\,.
\end{equation}

Although, in this case, there is no need to include the 
$\alpha\mathbf{J}^{1}$ term, we will keep it to illustrate its effects. Then,

\begin{equation}
  \mathbf{L}
  \doteq
  d\left[\tfrac{1}{2}e^{-a\phi} \star F\wedge A
    +\alpha\mathbf{J}^{1}
\right]\,,  
\end{equation}

\noindent
and, if $k$ generates a transformation that leaves invariant all the fields

\begin{equation}
  \begin{aligned}
    \imath_{k}\mathbf{L}
    & \doteq
      \imath_{k} d\left[
      \tfrac{1}{2}e^{-a\phi} \star F\wedge A
      +\alpha\mathbf{J}^{1}
      \right]
    \\
    & \\
    & =
      -d\imath_{k}\left[
      \tfrac{1}{2}e^{-a\phi} \star F\wedge A
      \right]
      +\delta_{\Lambda_{k}}\left[
      \tfrac{1}{2}e^{-a\phi} \star F\wedge A
      \right]
    \\
    & \\
    & \hspace{.5cm}
      +\alpha\left\{   -d\imath_{k}\mathbf{J}^{1}
      +\delta_{\Lambda_{k}}\mathbf{J}^{1}\right\}\,.
  \end{aligned}
\end{equation}

The first term gives

\begin{equation}
  \begin{aligned}
-d\imath_{k}\left[
\tfrac{1}{2}e^{-a\phi} \star F\wedge A
    \right]
    & =
        d\left[
     -\tfrac{1}{2}\tilde{P}_{k}F
      -\tfrac{1}{2}e^{-a\phi} \star F\imath_{k}A
      \right]\,,
  \end{aligned}
\end{equation}

\noindent
while the second gives

\begin{equation}
\delta_{\Lambda_{k}}\left[
  \tfrac{1}{2}e^{-a\phi} \star F\wedge A
\right]
  =
  d\left[\tfrac{1}{2}e^{-a\phi} \star F\chi_{k}\right]
      +\tfrac{1}{2}\mathbf{E}_{A}\chi_{k}\,,
\end{equation}

\noindent
where

\begin{equation}
  \chi_{k}
  =
  \imath_{k}A -P_{k}\,,
\end{equation}

\noindent
is the parameter of the compensating gauge transformation. Thus,

\begin{equation}
  \boldsymbol{\tau}(k,s)
  =
  \tfrac{1}{2}e^{-a\phi} \star F\chi_{k}=
  \tfrac{1}{2}e^{-a\phi} \star F(\imath_{k}A -P_{k})\,,
\end{equation}

\noindent
and the sum of the first two terms gives

\begin{equation}
d\left[
      -\tfrac{1}{2}\left(\tilde{P}_{k} F +P_{k}e^{-a\phi} \star F\right)
      \right]\,.  
\end{equation}

The third and fourth terms are \cite{Ballesteros:2023iqb} 

\begin{subequations}
  \begin{align}
    -d\imath_{k}\mathbf{J}^{1}
    +\delta_{\Lambda_{k}}\mathbf{J}^{1}
    & =
      d\mathbf{Q}_{k}+\tfrac{1}{2}a\mathbf{E}_{A}\chi_{k}\,,
      \\
    & \nonumber \\
    \mathbf{Q}_{k}
    & =
      -\imath_{k}\star d\phi
      -\tfrac{1}{2}a\left(\tilde{P}_{k} F +P_{k}e^{-a\phi} \star F\right)\,,
      \\
    & \nonumber \\
    d\mathbf{Q}_{k}
    & \doteq 
      0\,.
  \end{align}
\end{subequations}

$\mathbf{Q}_{k}$ is, therefore, a conserved 2-form charge. Its integral at
infinity gives the scalar charge of the dilaton
\cite{Pacilio:2018gom,Ballesteros:2023iqb}.

Combining these two results, we get

\begin{equation}
  \begin{aligned}
  \imath_{k}\mathbf{L}
    & \doteq
        d\left[
      -\tfrac{1}{2}\left(\tilde{P}_{k} F +P_{k}e^{-a\phi} \star F
      -\alpha\mathbf{Q}_{k}\right)
      \right]\,,
  \end{aligned}
\end{equation}

\noindent
which leads to the family of generalized generalized Komar charges

\begin{equation}
  \begin{aligned}
  \mathbf{K}[k]
  & =
  -\star (e^{a}\wedge e^{b}) P_{k\, ab}
  +\tfrac{1}{2}\left(P_{k}e^{-a\phi} \star F -\tilde{P}_{k} F\right)
    +\alpha\mathbf{Q}_{k}
    \\
    & \\
    & =
  -\star (e^{a}\wedge e^{b}) P_{k\, ab}
      +\tfrac{1}{2}\left[
      P_{k}(1-a\alpha)e^{-a\phi} \star F 
      -(1+a\alpha)\tilde{P}_{k} F \right]
    -\alpha\imath_{k}\star d\phi\,.
  \end{aligned}
\end{equation}

If (with the right boundary conditions for the momentum maps) the standard
Komar charge ($\alpha=0$) gives the mass when integrated at infinity, it is
clear that for $\alpha\neq 0$ the above charge will contain a term
proportional to the scalar charge of the dilaton and to $\alpha$. Furthermore,
for $\alpha\neq 0$, the terms involving the electric and magnetic momentum
maps and the Maxwell field strength are not invariant under discrete
electric-magnetic duality transformations, as it must because the equations of
motion are \cite{Mitsios:2021zrn}. Thus, it is reasonable to choose
$\alpha=0$. However, we must stress that the Smarr formula one gets is
independent of $\alpha$ if one uses the relation between the scalar charge,
the electric and magnetic potentials and charges
\cite{Pacilio:2018gom,Ballesteros:2023iqb} which follows from the fact
that $\mathbf{Q}_{k}$ satisfies its own, independent, Gauss law.

\section{ Example 2: minimal 5-dimensional supergravity}
\label{sec-example2N1D5}

The bosonic sector of minimal 5-dimensional supergravity contains an Abelian
gauge field, $V$, with field strength $G=dV$ coupled to gravity. Its action is

\begin{equation}
\label{eq:minimalN1d5action}
\begin{aligned}
  S[e^{a},V]
  & =
  \frac{1}{16\pi G_{N}^{(5)}} \int 
  \left[\star (e^{a}\wedge e^{b}) \wedge R_{ab}
    -\tfrac{1}{2}G\wedge \star G
        +\tfrac{1}{3^{3/2}} G\wedge G \wedge V
      \right]
        \\
  & \\
  & \equiv
  \int \mathbf{L}\,.
\end{aligned}
\end{equation}

A general variation of this action takes the form

\begin{equation}
  \label{eq:generalvariationd5}
  \delta S
  =
  \int \left\{
    \mathbf{E}_{a}\wedge \delta e^{a} +\mathbf{E}\wedge \delta V
    +d\mathbf{\Theta}(\varphi,\delta \varphi)
  \right\}\,,
\end{equation}

\noindent
where, ignoring for the moment the overall normalization factor of
$\left(16\pi G_{N}^{(5)}\right)^{-1}$,

\begin{subequations}
  \begin{align}
    \label{eq:Ea}
  \mathbf{E}_{a}
  & =
     \imath_{a}\star (e^{c}\wedge e^{d})\wedge R_{cd}
      +\tfrac{1}{2}\left(\imath_{a}G\wedge \star G
      -G\wedge \imath_{a}\star G
      \right)\,,
    \\
    & \nonumber \\
    \label{eq:E}
    \mathbf{E}
    & =
      -d\star G+\tfrac{1}{3^{1/2}}G\wedge G\,,
 \end{align}
 \end{subequations}

\noindent
are the equations of motion and

\begin{equation}
  \begin{aligned}
    \mathbf{\Theta}(\varphi,\delta\varphi)
    & =
    -\star (e^{a}\wedge e^{b})\wedge \delta \omega_{ab}
    +\left(\star G
    -\tfrac{2}{3^{3/2}}G\wedge V\right)\wedge \delta V\,,
  \end{aligned}
\end{equation}

\noindent
is the pre-symplectic 4-form of the theory.

Under the global rescaling transformations

\begin{equation}
  e^{\prime\, a}  =  \lambda e^{a}\,,
  \hspace{1cm}
  V' = \lambda V\,,
\end{equation}

\noindent
the action rescales with weight $3$ and

\begin{equation}
  \mathbf{L}
  =
  \tfrac{1}{3}\mathbf{E}_{a}\wedge e^{a}
  +\tfrac{1}{3}\mathbf{E}\wedge V +d\left(\tfrac{1}{3}\star G\wedge V\right)\,,
  \,\,\,\,\,\,
  \Rightarrow
  \,\,\,\,\,\,
  \mathbf{J}^{0}
  =
  \tfrac{1}{3}\star G\wedge V\,.
\end{equation}

This theory does not have any global symmetry. Thus, $\mathbf{J}^{0}$ is
defined unambiguously.

Under the gauge transformations

\begin{equation}
  \delta_{\chi}V
  =
  d\chi\,,
\end{equation}

\noindent
the action is invariant up to a total derivative with

\begin{equation}
  \mathbf{B}(\chi,\varphi)
  =
  -\tfrac{1}{3^{3/2}}G\wedge G\chi\,,
\end{equation}

\noindent
while $\mathbf{J}^{0}$ transforms as

\begin{equation}
    \delta_{\chi}\mathbf{J}^{0}
     =
      -\tfrac{1}{3}\mathbf{E}\chi -\mathbf{B}(\chi,\varphi)
      +d\boldsymbol{\tau}(\chi,\varphi)\,,
\end{equation}

\noindent
with

\begin{equation}
    \boldsymbol{\tau}(\chi,\varphi)
    =
      -\tfrac{1}{3}\star G\chi\,.      
\end{equation}

The Noether--Wald charge can be computed using the expressions discussed in
Section~\ref{sec-generalizedKomarcharges}. In particular, we can use
Eq.~(\ref{eq:dQ=ikL+B}) to compute it on-shell and for a Killing vector. Using
the definition of electric Eq.~(\ref{eq:Maxwellmomentummapequation}) and
Lorentz Eq.~(\ref{eq:Lorentzmomentummapequation}) momentum maps plus the
equations of motion and Bianchi identities one arrives at

\begin{equation}
  \mathbf{Q}[k]
  \doteqdot
  -\star(e^{a}\wedge e^{b})P_{k\, ab} +P_{k}\left(\star G
    -\tfrac{2}{3^{3/2}}G\wedge V\right)\,.
\end{equation}

On the other hand, using the magnetic momentum map equation
Eq.~(\ref{eq:magneticMaxwellmomentummapequationd5}) and the definition of the
compensating gauge parameter $\chi_{k}=\imath_{k}V-P_{k}$, we find

\begin{equation}
  -\imath_{k}\mathbf{J}^{0}
  +\boldsymbol{\tau}(\chi_{k},s)
  =
  \tfrac{1}{3}\tilde{P}_{k}\wedge G
  -\tfrac{2}{3^{3/2}} P_{k}G\wedge V
  +\tfrac{1}{3}P_{k}\star G\,,
\end{equation}

\noindent
so the Komar charge is given by these two fully equivalent expressions:

\begin{subequations}
  \begin{align}
    \mathbf{K}[k]
    & =
      \star(e^{a}\wedge e^{b})P_{k\, ab}
      -\tfrac{2}{3}P_{k}\star G
      +\tfrac{1}{3}\tilde{P}_{k}\wedge G\,,
    \\
    & \nonumber \\
    \mathbf{K}[k]
    & =
      \star(e^{a}\wedge e^{b})P_{k\, ab}
      -\tfrac{2}{3}P_{k}\left(\star G -\tfrac{1}{3^{1/2}}G\wedge V \right)
      +\tfrac{1}{3}\left(\tilde{P}_{k} -\tfrac{2}{3^{1/2}}P_{k}V\right)\wedge G\,,
  \end{align}
\end{subequations}

\noindent
the first of which is manifestly gauge invariant while the terms that enter
the second are products of the integrands of conserved charges and potentials
that satisfy generalized zeroth laws in black-hole spacetimes. This structure
was first found in Refs.~\cite{Elgood:2020mdx,Elgood:2020nls} is necessary to
derive Smarr formulas.

\section{Example: self-interacting scalar field}
\label{sec-example3selfinteractingscalar}

Our last example is that of a self-interacting scalar field $\phi$ coupled to
gravity in $d$ dimensions.\footnote{The 4-dimensional case was studied in
  Ref.~\cite{Ballesteros:2023muf}.} The action is

\begin{equation}
  \label{eq:actionselfcoupledscalars}
  \begin{aligned}
    S[e,\phi]
     & =
      \frac{1}{16\pi G_{N}^{(d)}}
      \int \left\{ (-1)^{d-1}\star(e^{a}\wedge e^{b})
      \wedge R_{ab}
      +\tfrac{1}{2}d\phi\wedge \star d\phi
       +(-1)^{d}\star V(\phi) \right\}
    \\
     & \\
     & \equiv
       \int \mathbf{L}\,.
  \end{aligned}
\end{equation}

The scalar potential must include at least one dimensionful coupling constant,
For the sake of simplicity we consider the case in which there is only one,
$g$, and

\begin{equation}
  V(\phi)
  =
  g^{2}f(\phi)\,.
\end{equation}

As conjectured in Ref.~\cite{Ortin:2021ade}, dimensionful coupling constants
play the role of thermodynamical variables in extended black-hole
thermodynamics. As shown in Ref.~\cite{Meessen:2022hcg}, the best way to find
the contribution of $g$ or any other coupling constant to the Smarr formula or
the first law of black-hole thermodynamics is to promote it to a scalar field
$g(x)$, adding at the same time a Lagrange-multiplier term depending on an
auxiliary $(d-1)$-form potential $C$ that forces it to be constant
on-shell. The action of this on-shell-equivalent theory is

\begin{equation}
  \label{eq:actiondilatonBHs2}
  \begin{aligned}
    S[e,\phi,g,C]
     & =
      \frac{1}{16\pi G_{N}^{(d)}}
      \int \left\{ -\star(e^{a}\wedge e^{b})
      \wedge R_{ab}
      +\tfrac{1}{2}d\phi\wedge \star d\phi
       +(-1)^{d}g^{2}\star f(\phi) +g^{2}dC\right\}
    \\
     & \\
     & \equiv
       \int \mathbf{L}\,.
  \end{aligned}
\end{equation}

Under an arbitrary infinitesimal variation of the fields

\begin{equation}
  \delta S
  =
  \int \left\{
    \mathbf{E}_{a}\wedge \delta e^{a} +\mathbf{E}_{\phi}\delta\phi
    +\mathbf{E}_{g}\delta g +\mathbf{E}_{C}\wedge \delta C
  +\mathbf{\Theta}(\varphi,\delta\varphi)\right\}\,,
\end{equation}

\noindent
where, ignoring the overall factor $(16\pi G_{N}^{(d)})^{-1}$, the equations
of motion and pre-symplectic potential are given by

\begin{subequations}
  \begin{align}
    \mathbf{E}_{a}
    & =
      \imath_{a}\star(e^{b}\wedge e^{c})\wedge R_{bc}
      +\tfrac{1}{2}\left(\imath_{a}d\phi\wedge \star d\phi
      +d\phi\wedge \imath_{a}\star d\phi \right) -\imath_{a}\star V\,,
    \\
    & \nonumber \\
    \mathbf{E}_{\phi}
    & =
      -d\star d\phi
      +(-1)^{d}\star g^{2} f'\,,
    \\
    & \nonumber \\
    \mathbf{E}_{g}
    & =
      2g\left(dC+(-1)^{d}\star f\right)\,,
    \\
    & \nonumber \\
    \mathbf{E}_{C}
    & =
      -2gdg\,,
    \\
    & \nonumber \\
    \mathbf{\Theta}(\varphi,\delta \varphi)
    & =
      -\star(e^{a}\wedge  e^{b})\wedge \delta \omega_{ab}
    +\star d\phi\delta\phi +g^{2}\delta C\,.
  \end{align}
\end{subequations}

This theory is exactly invariant under the gauge transformations

\begin{equation}
  \delta_{\chi} C
  =
  d\chi\,,
\end{equation}

\noindent
where $\chi$ is an arbitrary $(d-2)$-form.

Under the global rescaling\footnote{Notice that in the original theory $g$,
  being a constant that defines it, cannot be rescaled.}

\begin{equation}
  e^{\prime\, a}  =  \lambda e^{a}\,,
  \hspace{.5cm}
  g' = \lambda^{-1} g\,,
  \hspace{.5cm}
  C' = \lambda^{d} C\,,
\end{equation}

\noindent
the Lagrangian rescales with weight $(d-2)$ and, therefore, we can write

\begin{subequations}
  \begin{align}
  \mathbf{L}
  & =
  \frac{1}{(d-2)}\mathbf{E}_{a}\wedge e^{a}
  -\frac{1}{(d-2)}\mathbf{E}_{g} g +\frac{d}{d-2}\mathbf{E}_{C}\wedge C
    +d\mathbf{J}^{0}\,,
    \\
    & \nonumber \\
    \mathbf{J}^{0}
    & =
      \frac{d}{(d-2)}g^{2}C\,.
  \end{align}
\end{subequations}

We can immediately find

\begin{subequations}
  \begin{align}
    \boldsymbol{\tau}(\chi,\varphi)
    & =
      \frac{d}{(d-2)}g^{2}\chi\,,
    \\
    & \nonumber \\
  \imath_{k}\mathbf{J}^{0}
    & =
      \frac{d}{(d-2)}g^{2}\imath_{k}C\,,    
  \end{align}
\end{subequations}

\noindent
and using the momentum map equation

\begin{equation}
  \imath_{k}dC
  +dP_{k}
  =
  0\,,
\end{equation}

\noindent
that defines the momentum map $(d-2)$-form $P_{k}$, we find the Noether--Wald
$(d-2)$-form

\begin{equation}
  \mathcal{O}_{s}\imath_{k}\mathbf{L}
  =
  d\left[(-1)^{d}\star (e^{a}\wedge e^{b}) P_{k\, ab}
    -g^{2}P_{k}\right]
  =
  d\mathbf{Q}[k]\,.
\end{equation}

Combining these results, using the expression $\chi_{k}=\imath_{k}C-P_{k}$ for
the parameter of the compensating gauge transformation, we get the following
expression for the generalized Komar charge

\begin{equation}
  \mathbf{K}[k]
  =
  (-1)^{d-1}\star (e^{a}\wedge e^{b}) P_{k\, ab}
    -\frac{2}{d-2}g^{2}P_{k}\,.
\end{equation}

This result is equivalent on-shell to the result presented in
Ref.~\cite{Ballesteros:2023muf} in the 4-dimensional case but its form is
closer to the ``potential times conserved charge'' form of the terms we have
found in previous cases and which is necessary to derive Smarr formulas, or,
at least, to simplify their derivation.

\section{Conclusions}
\label{sec-conclusions}

In this paper we have shown how to derive construct systematically generalized
Komar charges of theories possessing a transformation of the fields that
rescales the action. The existence of this transformation and a field-theory
version of the Euler theorem allow us to write the on-shell Lagrangian as a
total derivative, which is one of the key steps in the construction of Komar
charges.

We have shown how the algorithm works in several examples. The last example
belonged to a non-linear theory and we have shown how one can still find the
needed transformation by promoting the coupling constant to a scalar
field. This trick can also be used in theories of non-linear electrodynamics
coupled to gravity \cite{kn:Barbagallo} and we are currently exploring other
interesting non-linear theories in which it may work.

In the main text we have stressed the fact that the global transformations of
the fields that rescale the action need not be rescalings of the fields. In
Ref.~\cite{Gomez-Fayren:2024cpl} a higher-form transformation that rescales
the Einstein--Hilbert action\footnote{This transformation was combined with an
  ordinary rescaling of the fields in order to obtain a symmetry.} with
Kaluza--Klein boundary conditions was found and it can also be used to rewrite
the on-shell Einstein--Hilbert Lagrangian as a total derivative. We will
consider this interesting case in a forthcoming publication
\cite{Barbagallo:2025fkg}.

\section*{Acknowledgments}

This work has been supported in part by the MCI, AEI, FEDER (UE) grants
PID2021-125700NB-C21 (``Gravity, Supergravity and Superstrings'' (GRASS)) and
IFT Centro de Excelencia Severo Ochoa CEX2020-001007-S.  The work of JLVC has
been supported by CSIC JAE-INTRO grant JAEINT-24-02806. TO wishes to thank
M.M.~Fern\'andez for her permanent support.

\appendix

\section{The magnetic momentum map in minimal $d=5$ supergravity}
\label{app:magneticmomentummap}

Assuming that the transformation $\delta_{k}$ leaves invariant all the fields,
since $\star G$ is a tensor

\begin{equation}
  0
  =
  \delta_{k}\star G
  =
  -\mathcal{L}_{k}\star G
  =
  -\left(\imath_{k}d +d\imath_{k}\right)\star G\,.
\end{equation}

Then, taking the interior product of $k$ with the equation of motion of the
1-form Eq.~(\ref{eq:E}) and using the above result and the definition of the
(``electric'') momentum map Eq.~(\ref{eq:Maxwellmomentummapequation}) with $F$
replaced by $G$

\begin{equation}
  \begin{aligned}
    \imath_{k}\mathbf{E}
    & =
      -\imath_{k}d\star G +\tfrac{2}{3^{1/2}}\imath_{k}G\wedge G
    \\
    & \\
    & =
      d\imath_{k}\star G -\tfrac{2}{3^{1/2}}dP_{k}\wedge G
    \\
    & \\
    & =
      d\left(\imath_{k}\star G -\tfrac{2}{3^{1/2}}P_{k}G\right)\,,
  \end{aligned}
\end{equation}

\noindent
where we have used the Bianchi identity $dG=0$ in the last step.

Thus, on-shell and locally there exists a 1-form $\tilde{P}_{k}$ (the
``magnetic'' momentum map) satisfying the equation

\begin{equation}
  \label{eq:magneticMaxwellmomentummapequationd5}
  \imath_{k}\star G -\tfrac{2}{3^{1/2}}P_{k}G +d\tilde{P}_{k}
  =
  0\,.
\end{equation}



\begin{thebibliography}{99}

\bibitem{Abbott:1981ff}
L.~F.~Abbott and S.~Deser,
``Stability of Gravity with a Cosmological Constant,''
Nucl. Phys. B \textbf{195} (1982), 76-96
\doi{10.1016/0550-3213(82)90049-9}
  
\bibitem{Barnich:2001jy}
G.~Barnich and F.~Brandt,
``Covariant theory of asymptotic symmetries,
conservation laws and central charges,''
Nucl. Phys. B \textbf{633} (2002), 3-82
\doi{10.1016/S0550-3213(02)00251-1}
[\hepth{0111246} [hep-th]].

\bibitem{Deser:2002jk}
S.~Deser and B.~Tekin,
``Energy in generic higher curvature gravity theories,''
Phys. Rev. D \textbf{67} (2003), 084009
\doi{10.1103/PhysRevD.67.084009}
[\hepth{0212292} [hep-th]].

\bibitem{Adami:2017phg}
H.~Adami, M.~R.~Setare, T.~C.~Sisman and B.~Tekin,
``Conserved Charges in Extended Theories of Gravity,''
Phys. Rept. \textbf{834} (2019), 1
\doi{10.1016/j.physrep.2019.08.003}
[\arxiv{1710.07252} [hep-th]].

\bibitem{Komar:1958wp}
A.~Komar,
``Covariant conservation laws in general relativity,''
Phys. Rev. \textbf{113} (1959), 934-936
\doi{10.1103/PhysRev.113.934}

\bibitem{Barnich:2003xg}
G.~Barnich,
``Boundary charges in gauge theories: Using Stokes theorem in the bulk,''
Class. Quant. Grav. \textbf{20} (2003), 3685-3698
\doi{10.1088/0264-9381/20/16/310}
[\hepth{0301039} [hep-th]].

\bibitem{Bardeen:1973gs}
J.~M.~Bardeen, B.~Carter and S.~W.~Hawking,
``The Four laws of black hole mechanics,''
Commun. Math. Phys. \textbf{31} (1973), 161-170
\doi{10.1007/BF01645742}

\bibitem{Carter:1973rla}
B.~Carter,
``Black holes equilibrium states,''
Contribution to: Les Houches Summer School of Theoretical Physics, 57-214.

\bibitem{Magnon:1985sc}
A.~Magnon,
``On Komar integrals in asymptotically anti-de Sitter space-times,''
J. Math. Phys. \textbf{26} (1985), 3112-3117
\doi{10.1063/1.526690}

\bibitem{Bazanski:1990qd}
S.~L.~Bazanski and P.~Zyla,
``A Gauss type law for gravity with a cosmological constant,''
Gen. Rel. Grav. \textbf{22} (1990), 379-387

\bibitem{Liberati:2015xcp}
S.~Liberati and C.~Pacilio,
``Smarr Formula for Lovelock Black Holes: a Lagrangian approach,''
Phys. Rev. D \textbf{93} (2016) no.8, 084044
\doi{10.1103/PhysRevD.93.084044}
[\arxiv{1511.05446} [gr-qc]].

\bibitem{Ortin:2021ade}
T.~Ort\'{\i}n,
``Komar integrals for theories of higher order
in the Riemann curvature and black-hole chemistry,''
JHEP \textbf{08} (2021), 023
\doi{10.1007/JHEP08(2021)023}
[\arxiv{2104.10717} [gr-qc]].

\bibitem{Mitsios:2021zrn}
D.~Mitsios, T.~Ort\'{\i}n and D.~Pere\~niguez,
``Komar integral and Smarr formula for axion-dilaton
black holes versus S duality,''
JHEP \textbf{08} (2021), 019
\doi{10.1007/JHEP08(2021)019}
[\arxiv{2106.07495} [hep-th]].

\bibitem{Meessen:2022hcg}
P.~Meessen, D.~Mitsios and T.~Ort\'{\i}n,
``Black hole chemistry, the cosmological constant
and the embedding tensor,''
JHEP \textbf{12} (2022), 155
\doi{10.1007/JHEP12(2022)155}
[\arxiv{2203.13588} [hep-th]].

\bibitem{Ortin:2022uxa}
T.~Ort\'{\i}n and D.~Pere\~niguez,
``Magnetic charges and Wald entropy,''
JHEP \textbf{11} (2022), 081
\doi{10.1007/JHEP11(2022)081}
[\arxiv{2207.12008} [hep-th]].

\bibitem{Ortin:2024mmg}
T.~Ort\'{\i}n and M.~Zatti,
``A note on the Noether-Wald and generalized Komar charges,''
[\arxiv{2411.10420} [gr-qc]].

\bibitem{Ballesteros:2023muf}
R.~Ballesteros and T.~Ort\'{\i}n,
``Hairy black holes, scalar charges and extended thermodynamics,''
Class. Quant. Grav. \textbf{41} (2024) no.5, 055007
\doi{10.1088/1361-6382/ad210a}
[\arxiv{2308.04994} [gr-qc]].

\bibitem{Ballesteros:2024prz}
R.~Ballesteros and T.~Ort\'{\i}n,
``Generalized Komar charges and Smarr formulas for black holes
and boson stars,''
SciPost Phys. Core \textbf{8} (2025), 038
\doi{10.21468/SciPostPhysCore.8.2.038}
[\arxiv{2409.08268} [gr-qc]].

\bibitem{Bandos:2023zbs}
I.~Bandos and T.~Ort\'{\i}n,
``Noether-Wald charge in supergravity: the fermionic contribution,''
JHEP \textbf{12} (2023), 095
\doi{10.1007/JHEP12(2023)095}
[\arxiv{2305.10617} [hep-th]].

\bibitem{Bandos:2024rit}
I.~Bandos, P.~Meessen and T.~Ort\'{\i}n,
``Komar charge of $ \mathcal{N} $ = 2 supergravity
and its superspace generalization,''
JHEP \textbf{05} (2025), 231
\doi{10.1007/JHEP05(2025)231}
[\arxiv{2412.18510} [hep-th]].

\bibitem{Ortin:2024emt}
T.~Ort\'{\i}n and M.~Zatti,
``On the thermodynamics of the black holes of the
Cano-Ruip\'erez 4-dimensional string effective action,''
JHEP \textbf{06} (2025), 026
\doi{10.1007/JHEP06(2025)026}
[\arxiv{2411.10417} [hep-th]].

\bibitem{Ortin:2015hya}
T.~Ort\'{\i}n,
``Gravity and Strings'', 2nd edition, 
Cambridge University Press, 2015.

\bibitem{Barbagallo:2025fkg}
  G.~Barbagallo, J.~L.~V.~Cerdeira, C.~G\'omez-Fayr\'en,
  P.~Meessen and T.~Ort\'{\i}n,
``On the generalized Komar charge of Kaluza-Klein theories
and higher-form symmetries,''
[\arxiv{2506.15615} [hep-th]].
  
\bibitem{Elgood:2020svt}
Z.~Elgood, P.~Meessen and T.~Ort\'{\i}n,
``The first law of black hole mechanics in the
Einstein-Maxwell theory revisited,''
JHEP \textbf{09} (2020), 026
\doi{10.1007/JHEP09(2020)026}
[\arxiv{2006.02792} [hep-th]].

\bibitem{Elgood:2020mdx}
Z.~Elgood, D.~Mitsios, T.~Ort\'{\i}n and D.~Pere\~n\'{\i}guez,
``The first law of heterotic stringy black hole mechanics
at zeroth order in \ensuremath{\alpha}',''
JHEP \textbf{07} (2021), 007
\doi{10.1007/JHEP07(2021)007}
[\arxiv{2012.13323} [hep-th]].

\bibitem{Elgood:2020nls}
Z.~Elgood, T.~Ort\'{\i}n and D.~Pere\~n\'{\i}guez,
``The first law and Wald entropy formula of heterotic
stringy black holes at first order in $\alpha'$,''
JHEP \textbf{05} (2021), 110
\doi{10.1007/JHEP05(2021)110}
[\arxiv{2012.14892} [hep-th]].

\bibitem{Prabhu:2015vua}
K.~Prabhu,
``The First Law of Black Hole Mechanics for Fields
with Internal Gauge Freedom,''
Class. Quant. Grav. \textbf{34} (2017) no.3, 035011
\doi{10.1088/1361-6382/aa536b}
[\arxiv{1511.00388} [gr-qc]].

\bibitem{Ballesteros:2023iqb}
R.~Ballesteros, C.~G\'omez-Fayr\'en, T.~Ort\'{\i}n and M.~Zatti,
``On scalar charges and black hole thermodynamics,''
JHEP \textbf{05} (2023), 158
\doi{10.1007/JHEP05(2023)158}
[\arxiv{2302.11630} [hep-th]].

\bibitem{Pacilio:2018gom}
C.~Pacilio,
``Scalar charge of black holes in Einstein-Maxwell-dilaton theory,''
Phys. Rev. D \textbf{98} (2018) no.6, 064055
\doi{10.1103/PhysRevD.98.064055}
[\arxiv{1806.10238} [gr-qc]].

\bibitem{kn:Barbagallo}
  G.~Barbagallo and T.~Ort\'{\i}n,
  (in preparation).

\bibitem{Gomez-Fayren:2024cpl}
C.~G\'omez-Fayr\'en, T.~Ort\'{\i}n and M.~Zatti,
``Gravitational higher-form symmetries and the
origin of hidden symmetries in Kaluza-Klein compactifications,''
SciPost Phys. Core \textbf{8} (2025), 010
\doi{10.21468/SciPostPhysCore.8.1.010}
[\arxiv{2405.16706} [hep-th]].


\end{thebibliography}
\end{document}